\documentclass[a4paper,UKenglish, cleveref, autoref]{lipics-v2019}
\nolinenumbers
\title{Enhanced Performance and Privacy via Resolver-Less DNS} %TODO Please add

\titlerunning{Resolver-Less DNS}%optional, please use if title is longer than one line

\author{Erik Sy}{University of Hamburg}{}{}{}

\authorrunning{Erik Sy}%TODO mandatory. First: Use abbreviated first/middle names. Second (only in severe cases): Use first author plus 'et al.'

\Copyright{Erik Sy}%TODO mandatory, please use full first names. LIPIcs license is "CC-BY";  http://creativecommons.org/licenses/by/3.0/

%TODO mandatory: Please choose ACM 2012 classifications from https://dl.acm.org/ccs/ccs_flat.cfm 

%TODO mandatory; please add comma-separated list of keywords

\category{}%optional, e.g. invited paper

\relatedversion{}%optional, e.g. full version hosted on arXiv, HAL, or other respository/website
\supplement{}%optional, e.g. related research data, source code, ... hosted on a repository like zenodo, figshare, GitHub, ...

\hideLIPIcs  %uncomment to remove references to LIPIcs series (logo, DOI, ...), e.g. when preparing a pre-final version to be uploaded to arXiv or another public repository

\EventEditors{}
\EventNoEds{0}
\EventLongTitle{Conference on Principles of Distributed Systems (OPODIS 2019)}
\EventShortTitle{OPODIS 2019}
\EventAcronym{OPODIS}
\EventYear{2019}
\EventDate{December 17--19, 2016}
\EventLocation{Neuchâtel, Switzerland}
\EventLogo{}
\SeriesVolume{1}
\ArticleNo{1}
\begin{document}

\maketitle

\begin{abstract}
The domain name resolution into IP addresses can significantly delay connection establishments on the web.
Moreover, the common use of recursive DNS resolvers presents a privacy risk as they can closely monitor the user's browsing activities. 
In this paper, we present a novel HTTP response header allowing web server to provide their clients with relevant DNS records.
Our results indicate, that this resolver-less DNS mechanism allows user agents to save the DNS lookup time for subsequent connection establishments.
We find, that this proposal saves at least 80~ms per DNS lookup for the one percent of users having the longest round-trip times towards their recursive resolver.
Furthermore, our proposal decreases the number of DNS lookups and thus improves the privacy posture of the user towards the used recursive resolver.
Comparing the security guarantees of the traditional DNS to our proposal, we find that resolver-less DNS achieves at least the same security properties.
In detail, it even improves the user's resilience against censorship through tampered DNS resolvers.  
\end{abstract}

\section{Introduction}
The Internet community aims to improve the performance of connection establishments on the web with means such as the QUIC transport protocol and TLS~1.3 zero Round-Trip Time (0-RTT) handshakes.
However, many connection establishments on the web require a prior resolution of the domain name into IP addresses.
The Domain Name System (DNS) is responsible for this task.
Thus, DNS lookups can become a performance bottleneck on the web.
Reducing the time required to resolve domain names into IP addresses benefits the page load time of websites.
Following this, it leads to an improved user experience during web browsing~\cite{Varvello:2016:EPC:2999572.2999590} and increases the per-user revenue of online service provider~\cite{velocity}.

To further improve the performance of resolving domain names into IP addresses, we propose a novel HTTP response header.
This new header field \textit{DNS-Record} allows web server to provide their clients the resolved IP addresses for relevant domain names.
Subsequently, the client can directly begin with the connection establishment saving the DNS lookup time.
Furthermore, this approach improves user privacy as only a smaller fraction of a client's browsing activities can be observed by the used DNS resolver.

In summary, this paper makes the following contributions:
\begin{itemize}
\item We propose resolver-less DNS allowing web servers to push DNS records to their clients via HTTP response header.

\item We demonstrate the performance improvements gained by our proposal and investigate the resulting DNS overhead for the web server. Our results indicate that our proposal allows clients to save the DNS lookup time. We find, that the DNS lookup accounts for at least 80~ms for the one percent of users having the longest round-trip times towards their DNS resolver.

\item We review the privacy and security impact of our proposal. We observe that resolver-less DNS significantly improves the user's privacy posture towards its recursive DNS resolver. Furthermore, we find that our proposed design provides at least the security guarantees of the traditional DNS\@.

\end{itemize}

The remainder of this paper is structured as follows: Section~\ref{sec:Problem} provides an overview of the traditional DNS and introduces its performance, privacy and security problems that we aim to mitigate.
Section~\ref{sec:Design} summarizes the proposed resolver-less DNS design.
Evaluation results and a discussion of our proposal are presented in Section~\ref{sec:Evaluation} and~\ref{sec:Discussion}. 
Related work is reviewed in Section~\ref{sec:Related}.
Section~\ref{sec:Conclusion} concludes the paper.

\section{DNS Overview and Problem Statement}~\label{sec:Problem}
In this section, we first describe a popular DNS deployment using recursive DNS resolvers.
Subsequently, we review performance, privacy and security problems of this approach to resolve domain names into IP addresses.

\subsection{DNS Overview}

\begin{figure}[t]
\centering
\includegraphics[width=0.47 \textwidth]{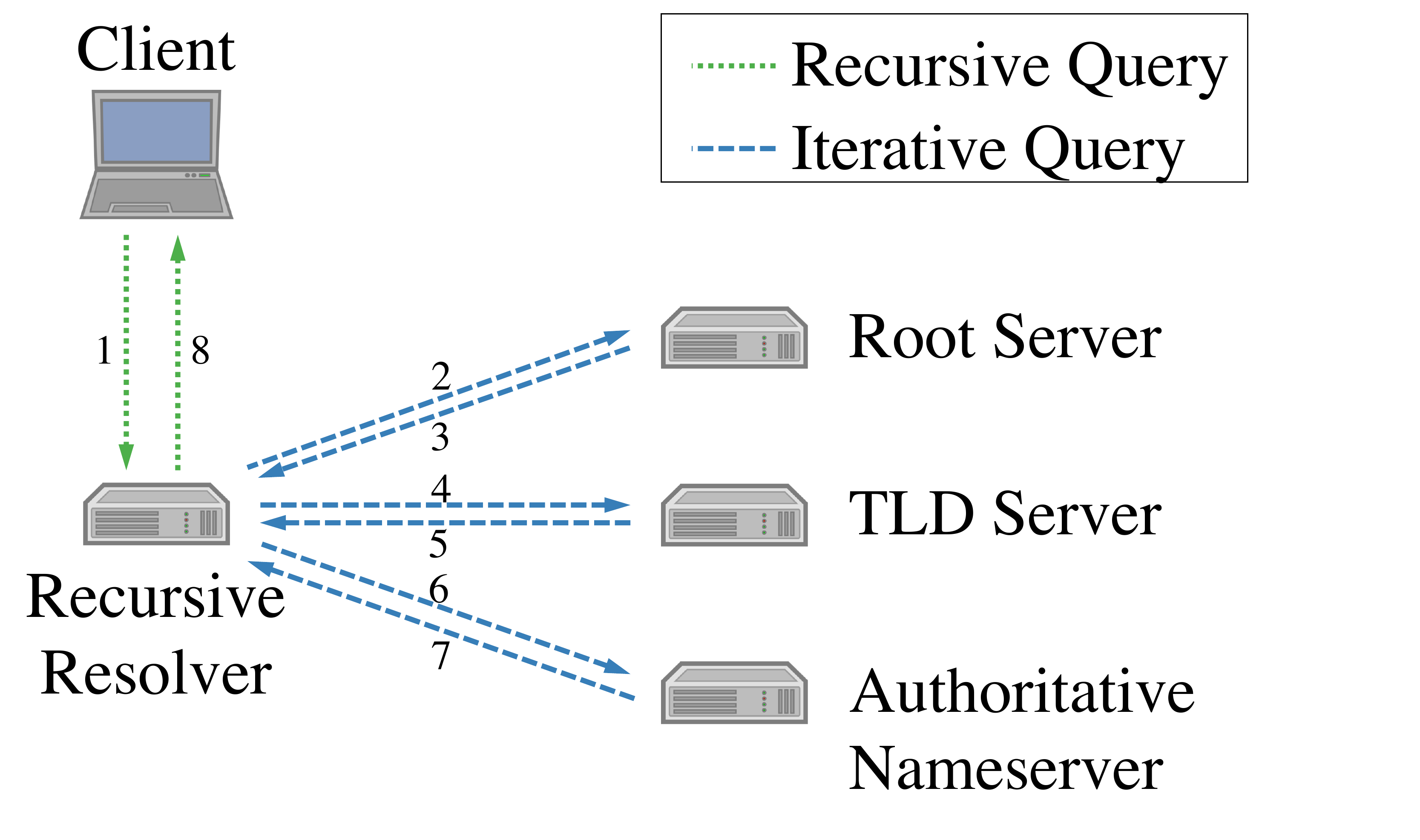}
\caption{Schematic of a DNS lookup using a recursive resolver. The numbers indicate the sequence of the data flows.}
\label{fig:Recursive_DNS}
%\vskip -12pt
\end{figure}

The Domain Name System (DNS) is responsible for resolving domain names into IP addresses and presents an essential component of our Internet infrastructure.
Real-world DNS traffic indicates that median users conduct about 1384 DNS lookups per day to 372 different hostnames~\cite{herrmann2013behavior}.
Popular operating systems and web browsers have a local DNS cache to reduce the number of required DNS lookups.
In total, between 12.9\% and 20.4\% of the average user's TCP connections directly follow a DNS query to the recursive resolver~\cite{jung2002dns}.
Figure~\ref{fig:Recursive_DNS} provides a schematic of a DNS lookup using a recursive resolver.
DNS recursive resolvers are usually provided by the Internet Service Provider (ISP) or the client may use a publicly accessible recursive resolver such as Google DNS\@.
The DNS lookup starts with the user's query containing the domain name it aims to resolve, shown as arrow~1 in Figure~\ref{fig:Recursive_DNS}.
Upon receiving this query, the recursive resolver can either directly resolve the domain name using its DNS cache or it experiences a cache miss.
In case of a cache miss, the recursive resolver investigates the authoritative nameserver of the respective domain name.
This can be done via an iterative query along the DNS hierarchy involving the DNS root server and the Top Level Domain (TLD) server, as can be seen in Figure~\ref{fig:Recursive_DNS}.
Finally, the recursive resolver queries the authoritative nameserver to resolve the respective domain name into an IP address.
Receiving this response from the authoritative nameserver, the recursive resolver forwards it towards the client as indicated by arrow 8 in Figure~\ref{fig:Recursive_DNS}.
Note, that a recursive resolver may cache information about TLD servers and authoritative nameservers allowing it to skip steps of this complete query flow in future queries.
For recursive resolvers cache hit rates larger than 80\% are reported~\cite{jung2002dns}.
Furthermore, studies suggest that recursive resolvers can often perform faster iterative queries compared to most of their users due to their advantageous position in the network topology~\cite{2019QuicSocks}.
Thus, it is more common for clients to resolve domain names into IP addresses using recursive resolvers compared to conducting the iterative queries themselves.  

\subsection{Problem Statement}

In this section, we describe performance, privacy and security problems of the traditional DNS that we aim to solve with our proposal. 

\subsubsection{Performance Problem}

A major motivation for clients to conduct DNS lookups is that they aim to send network packets towards the resolved IP address.
In these cases, the DNS lookup is an intermediate step delaying the start of the communication between client and server.
The DNS lookup time depends amongst others on the round-trip time between client and recursive resolver.
This round-trip time varies depending on the access network between a few milliseconds for wired connections and more than a hundred milliseconds for cellular 3G networks~\cite{hounsel2019analyzing}.
Another contribution to the lookup time presents the time required by the resolver to respond to the request.
For DNS cache hits these transactions complete in under 1~ms.
However, about 25\% of the average DNS transactions require between 10~ms and 1~second to complete~\cite{callahan2013modern}.
In total, the DNS lookup time can significantly increase the page loading time and therefore reduce the user-perceived quality of web experience~\cite{pan2003overview}.

\subsubsection{Privacy Problem}

There are several privacy risks associated with traditional DNS\@.
Here, we are focusing on the problems related to the recursive resolver.
Resolvers are in a privileged position that allows them to reconstruct a significant fraction of the user's browsing session based on the observed queries.
Furthermore, patterns within the observed DNS queries allow the recursive DNS resolver to identify the same user across several sessions with high probability~\cite{herrmann2013behavior}.
A study of the privacy policies of popular DNS resolvers indicates that many resolvers collect users' DNS traffic and subsequently monetize this data via advertising~\cite{Privacy_by_Infrastructure}.
Users are often unaware of the privacy risks associated with using a recursive resolver~\cite{kang2015my}.
Thus, an improved DNS would restrict the impact of recursive resolvers on user privacy.

\subsubsection{Security Problem}

In traditional DNS, the client has difficulties to detect that a recursive resolver modified the requested DNS records or even censored its response.
Measurements on DNS manipulations by recursive resolvers indicate that this presents a real-world problem enabling Internet censorship~\cite{pearce2017global}.
As a consequence, the censored online resources become unavailable to users.
To protect the user's ability to access online resources, an improved DNS should be more resilient against censorship.
Note, that a DNS security extension called \textit{DNSSEC} exists aiming to detect manipulations of DNS records.
However, \textit{DNSSEC} faces several deployment issues including a significant share of DNSSEC-enabled domains providing an incomplete or incorrect record preventing a validation~\cite{chung2017longitudinal}.
Furthermore, the used DNSSEC Key Signing Keys (KSKs) are often weak and the majority of DNSSEC-enabled domains does not rotate these KSKs~\cite{chung2017longitudinal}.
Thus, DNSSEC does not represent a feasible solution to the described security problem.

\section{Resolver-less DNS}\label{sec:Design}
In this section, we introduce the resolver-less DNS design.
This novel approach allows web servers to provide DNS records to their clients to accelerate their page loading time.
First, we summarize our design goals, before we present resolver-less DNS\@.

\subsection{Design Goals}
We aim to develop an approach to resolve hostnames into IP addresses that supports the following goals:

\begin{enumerate}
\item Deployable on today's Internet which excludes approaches requiring changes to middle-boxes, kernels of client machines, or the DNS protocol. 
\item Reduces the time required for the client to resolve a hostname and thus improves the webpage loading time.
\item Does not require additional Internet infrastructure such as proxies or dedicated DNS resolvers to be deployed.
\item Reduces the number of DNS queries sent to the DNS resolver. Thus, the resolver observes a diminished fraction of the user's browsing activities improving the privacy posture of the user.
\item Supports the performance-optimization EDNS client subnet.
\item Supports clients behind Network Address Translators (NAT).
\item Guarantees confidentiality accordingly to the client's connection with the web server providing the DNS records.
\item Security assurances are not weaker than using a recursive DNS resolver.
\item Provides an improved resilience to the threat of Internet censorship based on DNS\@.
\end{enumerate}

\subsection{Design}

To describe the design of the resolver-less DNS, we start by introducing a novel HTTP response header field of the type \textit{DNS-Record}.
Then, we summarize changes required to the client and server behavior to support the resolver-less DNS proposal.

\subsubsection{The \textit{DNS-Record} HTTP Response Header Field}
Our design introduces a novel HTTP response header field with the name \textit{DNS-Record}.
This header field contains the following key-value pairs:
\begin{itemize}
\item \textit{hostname} indicating the record's hostname,
\item \textit{A} providing the 32-bit IPv4 address of the hostname,
\item \textit{A\_TTL} providing the time-to-live in seconds of the A address record,
\item \textit{AAAA} providing the 128-bit IPv6 address of the hostname,
\item \textit{AAAA\_TTL} providing the time-to-live in seconds of the AAAA address record.
\end{itemize}
An example response header can look like this: \textit{DNS-Record: hostname=a.com; A=1.1.1.1; A\_TTL=299}.
To provide the client with a list of DNS records within the same HTTP response, the server can append further DNS records using a comma for separation.
Note, that HTTP provides a \textit{Link} response header field allowing the server to hint resources to the user agent that should be preloaded~\cite{rfc8288}.
Thus, our proposal extends this existing HTTP functionality to hint resources by providing additionally the corresponding DNS records.    

\subsubsection{Client Behavior}

A user agent such as a web browser receiving the \textit{DNS-Record} HTTP response header field saves the content within its DNS cache and marks this DNS record as retrieved via resolver-less DNS\@.
Subsequently, when the user agent intends to connect to this hostname, it retrieves the DNS record from its cache and validates that the record has not yet expired.
The client can use these DNS records to directly connect to the hostname without conducting a DNS lookup.
However, the client has usually no mechanism available to validate the integrity of these DNS records~\cite{chung2017longitudinal}.
To protect against web server serving malicious DNS records, we restrict the usage of resolver-less DNS records as follows.

\begin{figure}[t]
\centering
\includegraphics[width=0.90 \textwidth]{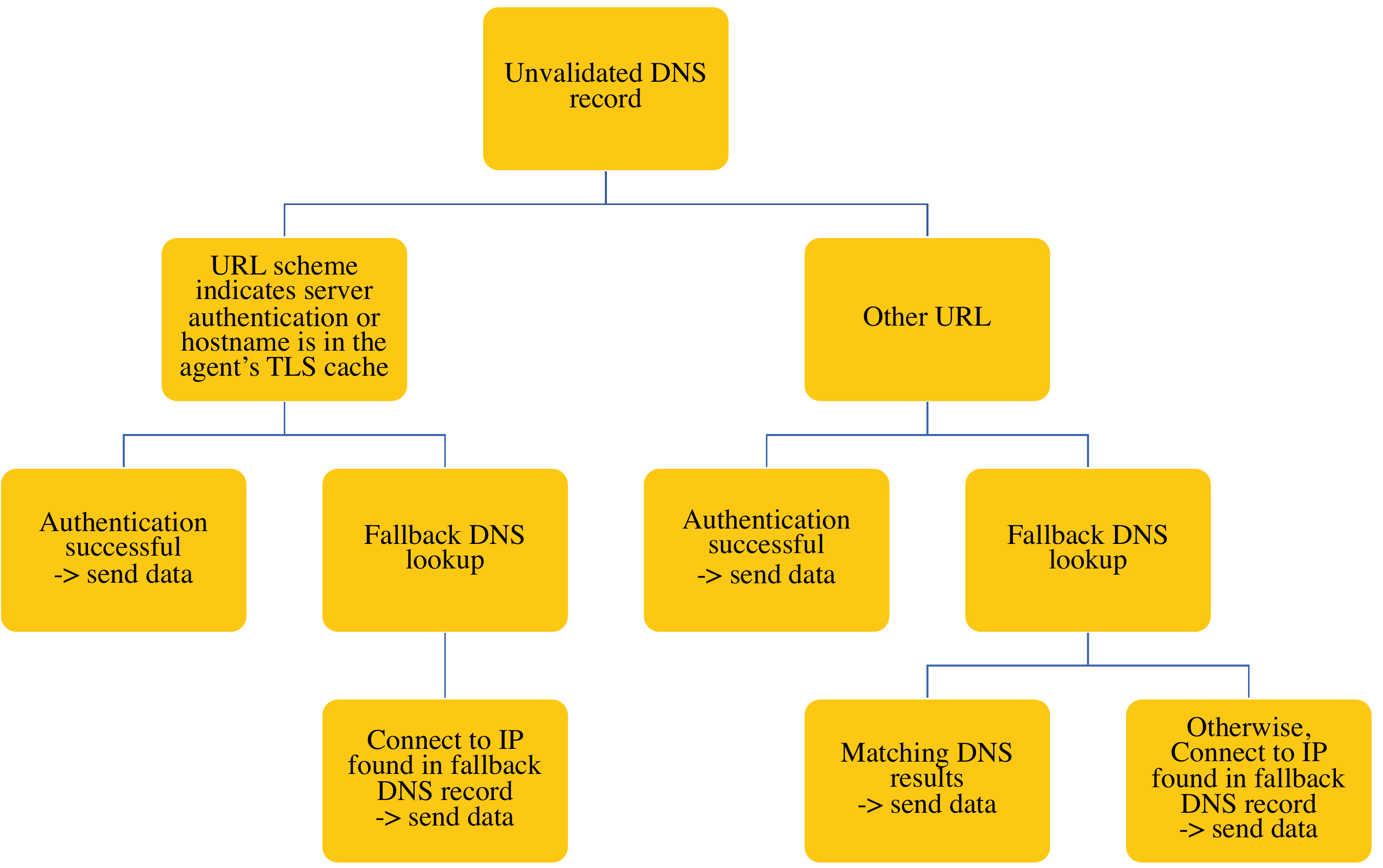}
\caption{Overview of the validation strategy for unvalidated DNS records.}
\label{fig:Resolver-less_Validation}
%\vskip -12pt
\end{figure}

First, clients are only allowed to send application data over an established connection if they validated the server's identity via a server authentication mechanism or a fallback DNS lookup.
Figure~\ref{fig:Resolver-less_Validation} presents the validation strategy applied by user agents on DNS records retrieved via resolver-less DNS.
Following this validation strategy, the user agent is allowed to directly start the connection establishment of protocols such as TCP, QUIC, and TLS based on the IP address in the unvalidated DNS record.
However, the client must only send application data over these connections, if it is encrypted in a manner that only the correct server entity is capable to decrypt this data.
TLS session resumption mechanisms~\cite{Sy_TLS_Tracking} are an example for these permitted encryption schemes.
On the left branch of the validation strategy, the user agent has an indication that the server can authenticate itself by providing an identity proof.
For example, the hostname has a record in the user agent's TLS cache or the scheme of the Uniform Resource Locator (URL) indicates server authentication as the scheme \textit{https} does.
Subsequently, the client attempts to authenticate the server's identity.
If the authentication is successful, this validates the used DNS record and the client can send application data over this connection.
Otherwise, the client needs to conduct a fallback DNS lookup to query the correct IP address of the hostname.
Then, the client connects and sends application data to the server at the IP address presented in the fallback DNS record.
Note, that the web is moving towards an encrypted ecosystem where today about 80\% of a user's HTTP requests use the secure HTTPS variant~\cite{HTTP_Archive}.
Furthermore, the upcoming HTTP version 3 uses always server authentication and transport encryption to exchange application data between client and server~\cite{ietf-quic-http-22}.
On the right branch, the user agent has no indication that the server is capable to authenticate itself.
However, in case the server authenticates it's identity during the connection establishment, then the client can directly send application data.
Otherwise, the client is required to conduct a fallback DNS lookup on the corresponding hostname.
In case of matching results between the fallback and the resolver-less DNS record, the established connection can be used to transmit application data.
Otherwise, the client uses the address within the fallback DNS record to establish a new connection and exchange application data.
Note, that a privacy versus performance tradeoff exists concerning the moment when the client starts the fallback DNS lookup.
The best performance is achieved, when the client always directly starts to validate the resolver-less DNS record via a fallback DNS lookup.
However, the client achieves the best privacy protection if the user agent only conducts necessary fallback DNS lookup based on the validation strategy.
In the following, we assume that the user agent conducts the fallback DNS lookup on the left branch of the validation strategy only when it becomes necessary.
However, the user agent will conduct the fallback DNS lookup on the right branch of the validation strategy as early as possible.

Second, there is the risk of web servers providing false DNS records leading to a negative performance impact when the user agent visits other websites.
As a result, we recommend user agents to restrict the contexts in which they trust the DNS records retrieved via resolver-less DNS\@.
For example, web browsers can restrict the usage of such DNS records within the context of the same browser tab or the same website.
In case of a context switch leading to reduced trust in these DNS records, we recommend a similar practice as described above.
In detail, the user agent starts the connection establishment towards the IP address found in the resolver-less DNS record and validates this information via a fallback DNS lookup.
For matching IP addresses within both DNS records, the user agent is permitted to send application data over the established connection.
Otherwise, the client prefers the name resolution of its fallback DNS mechanisms and uses a connection established to this IP address to exchange application data.

It seems reasonable to assume that websites provided by Google yield the same records as a DNS lookup to Google Public DNS\@.
Similar, the DNS records of websites served by Cloudflare can be trusted as much as a lookup to Cloudflare's DNS service.
Thus, we recommend to include a whitelist mechanism within user agents that does not apply the described restrictions if the client trusts a websites to provide authentic DNS records.
As Google and Cloudflare use a dedicated certificate authority to issue certificates for their hosted websites, this information could be used to associate websites to certain whitelisted organizations.

\subsubsection{Server Behavior}

Websites are primarily composed of HTML documents.
During a website visit, the user agent retrieves these documents along with other content such as images or videos from a web server using HTTP\@.
In the following, we use the term application server to refer to the application building the HTML documents.
The HTTPS server is tasked to communicate with the client and aims to serve whenever possible cached content to achieve a scalable system design.
Thus, the HTTPS server occasionally forwards requests to the application server to fetch updated versions of the cached content or to retrieve dynamic content directly from the application.

\begin{figure}[t]
\centering
\includegraphics[width=0.47 \textwidth]{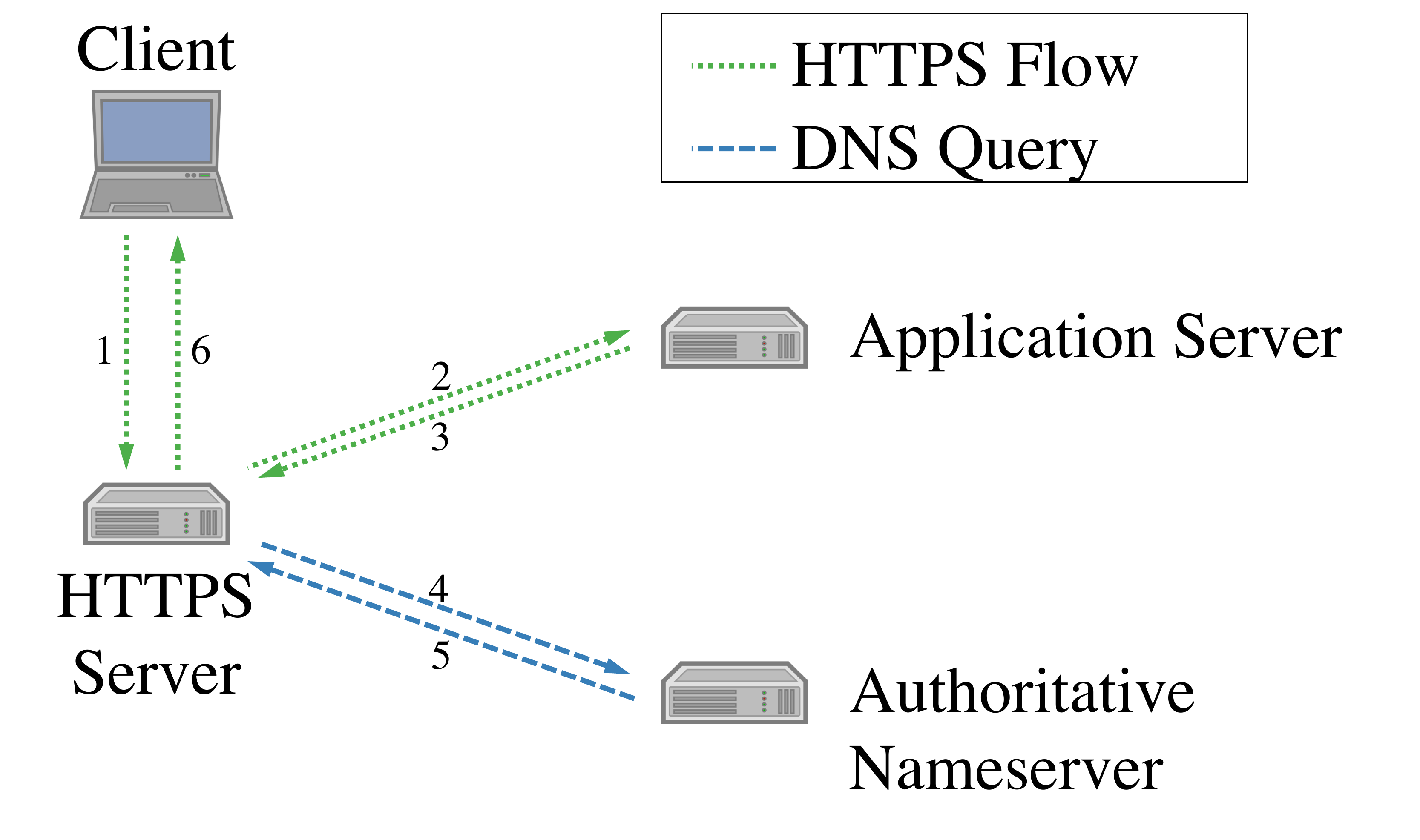}
\caption{Overview on the resolver-less DNS design where the HTTPS server provides the DNS record.}
\label{fig:Resolver-less_DNS}
%\vskip -12pt
\end{figure}

In the proposed design of a resolver-less DNS, we assign the task of DNS resolution towards the HTTPS server.
For example, if a website is served via a Content Delivery Network (CDN), it would be their task to provide the name resolution.
However, we want the application server to indicate the relevant hostnames as this prevents the HTTPS server from parsing HTML documents to learn the relevant hostnames.
We assume, that the HTTPS server and the application server communicate via HTTP with each other and the HTTP response header provided by the application will be merged into the server's HTTPS response to the client.
Thus, the application server could parse the HTML resource hints~\cite{Resource_Hints} such as instructions for DNS prefetching and writes for relevant hostnames \textit{DNS-Records} without resolution results into the HTTP response header.
Upon receiving such responses in step three of Figure~\ref{fig:Resolver-less_DNS}, the HTTPS server parses the HTTP response header and recognizes the necessity to include additional information into the \textit{DNS-Record} header field.

Subsequently, the HTTPS server conducts a DNS lookup of the respective hostnames and uses the retrieved DNS records to complete the \textit{DNS-Record} header field of the response.
The response can now be forwarded to the client as indicated by step 6 of Figure~\ref{fig:Resolver-less_DNS}.
Furthermore, the HTTPS server can now cache the response including the \textit{DNS-Record} header fields.
Upon serving the cached response to other clients, the server must dynamically recompute the remaining time-to-live for these \textit{DNS-Record} header fields.
Moreover, the HTTPS server should refresh the DNS records in the cached HTTP response before they expire.

Note, that HTTPS server can also decide to remove \textit{DNS-Record} header fields for some relevant hostnames from their response sent to the client.
Reasons for this practice include for example, HTTPS server facing temporary resource constraints may want to deactivate resolver-less DNS during these periods.
Furthermore, HTTPS server may not want to forward DNS records of relevant hostnames with a very short expiration time such as a few seconds. 

Authoritative nameserver may want to provide a different DNS response to users within different locations.
This approach allows nameserver to chose the IP address of a server having a rather low network latency towards the client.
To support the client in communicating with an advantageous server, the HTTPS server can support EDNS client subnet (ECS)~\cite{rfc7871}.
Here, the HTTPS server queries the nameserver with a truncated IP address of the client and receives a DNS record valid for a scope of IP addresses.
The HTTPS server can cache this DNS record and serve all clients matching the defined scope of IP addresses the same DNS record similar as described above.
In case the client's source address does not match the scope of the cached DNS record, the HTTPS server is required to conduct another DNS lookup with the client subnet information to retrieve an adequate DNS record. 

\section{Performance Evaluation}\label{sec:Evaluation}

In this section, we approximate the performance impact of the proposed resolver-less DNS\@. 
To begin with, we present a measurement of the round-trip time between typical clients and their resolvers.
This RTT presents a lower boundary for the time required by a client to conduct DNS lookups.
Subsequently, we investigate the expiration time of DNS records of popular hostnames and support for EDNS client subnet because these attributes impact the number of required DNS lookups by the web server.

\subsection{Lower Boundary of the DNS Lookup Time}

As the main benefit of our proposal, it allows clients to save the DNS lookup time and therefore reduces the webpage loading time.
To approximate this performance enhancement for real-world clients, we investigate the round-trip time between clients and their pre-configured DNS resolver.
Note, that this RTT presents only a lower boundary for the DNS lookup time as we neglect the time required by the resolver to process the query and to conduct possibly necessary iterative queries to other servers to provide the client with its response.

\subsubsection{Data Collection}

To obtain the real-world RTT between a broad range of clients and their pre-configured recursive resolvers, we make use of the RIPE Atlas network~\cite{ripe}.
The nodes of this RIPE Atlas network are distributed across several autonomous systems using various access technologies.
Using this network, we can task nodes to conduct custom DNS queries or ping measurements.
For this measurement, we selected a set of 800 RIPE Atlas nodes located in Germany.
Thus, we aim to obtain a realistic representation of typical Internet accesses in countries similar to Germany concerning their infrastructure. 

To measure the corresponding RTTs from different selected nodes, we use these nodes to ping their pre-configured recursive DNS resolver.
To ensure that resolvers using anycast services always return the same physical server and provide consistent RTTs, we start by investigating the IP address of the used recursive resolvers.
For this purpose, we control an authoritative nameserver for a subdomain similar to dnstest.a.com.
Subsequently, we send a query for a random subdomain in our DNS zone to the pre-configured recursive resolver such as random.dnstest.a.com.
Concurrently, we capture the network traffic of our authoritative nameserver and observe a DNS query for exactly the subdomain random.dnstest.a.com.
Thus, we reason that the sender's IP address of the observed DNS query in our network traffic is resolving the client's DNS query and is therefore associated with the node's recursive DNS resolver.
In cases, where this observed IP address does not match the locally configured DNS resolver within the node, we assume that these both IP addresses are colocated and yield about the same RTT\@.

We conducted our data collection on the 13th of June 2019 with five ping measurements between each node and its pre-configured DNS resolvers.
We computed the RTT as the average of these five ping measurements and obtained in total results for 650 nodes.
The main cause for failures can be attributed to resolvers not responding to our ping measurements.
Additionally, we observed also a small number of nodes experiencing failures during their DNS measurements.

Furthermore, we derived the autonomous system numbers of the IP addresses of all nodes and their resolvers.
For matching autonomous systems (AS) between the node and its recursive resolver, we assume that the resolver is provided by the node's Internet Service Provider.
This approach allows us to compare the performance of resolvers within the same AS versus the performance of resolvers belonging to a different AS as it would be likely for public DNS resolvers such as Goggle DNS\@.
In total, our data collection successfully obtained measurements from 650 nodes in Germany of which 474 nodes had at least one pre-configured DNS resolver within the same AS\@. 
298 of the investigated RIPE Atlas nodes had at least one pre-configured DNS resolver not matching their own AS\@.

\subsubsection{Results}

The traditional DNS has a performance overhead compared to our proposal as it requires clients to retrieve DNS records from their resolvers.
In this section, we evaluate the round-trip time between our 650 test nodes and their pre-configured DNS resolver.
This round-trip time represents the lower boundary of the performance overhead of the traditional DNS\@.
The round-trip time is a good approximation of the time required to conduct the DNS lookup if the resolver has a cached response for the client's query.
However, in case of a cache miss, the DNS lookup time can be significantly longer as the resolver conducts additionally iterative queries to the authoritative nameserver of the queried hostname.
Thus, a fraction of about 25\% of real-world DNS lookups require between 10~ms and 1~second to complete~\cite{callahan2013modern}.

\begin{figure}[t]
\centering
\includegraphics[width=0.60 \textwidth]{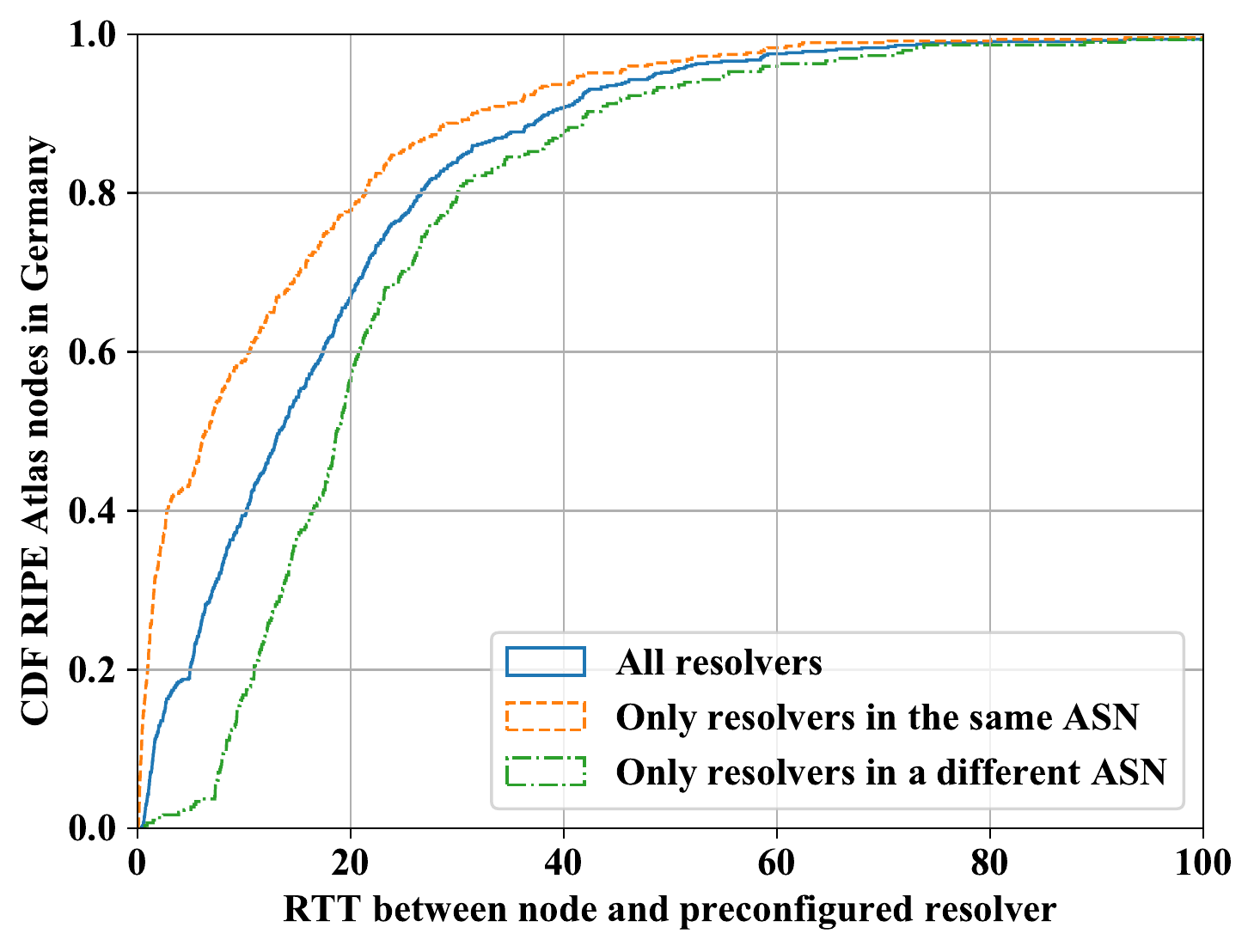}
\caption{Cumulative distribution of the Round-Trip Time (RTT) between the pre-configured DNS resolver and 650 RIPE Atlas nodes located in Germany. }
\label{fig:multi_cdf_dns}
%\vskip -12pt
\end{figure}

Figure~\ref{fig:multi_cdf_dns} plots a cumulative distribution of the round-trip time between 650 RIPE Atlas nodes located in Germany and their pre-configured DNS resolvers.
The solid, blue plot shows the corresponding distribution taking all DNS resolvers into account.
The dashed, orange plot focuses on a subset of DNS resolvers that match the ASN of our test node conducting the DNS lookup.
Here, we assume that this DNS resolver is provided by the client's ISP\@.
Whereas the dash-dotted, green plot indicates DNS lookups where the client and resolver are located in different autonomous systems as it can be expected for client's using a public DNS resolver.
Our results indicate that the investigated round-trip times are significantly faster when the client and the used DNS resolver share the same AS\@.
For example, 59\% of the tested nodes experienced less than 10~ms round-trip time when the resolver was located within the same AS\@.
However, only 17\% of the clients completed the round-trip within this duration when the peers did not share the same AS\@.
Moreover, we observe that a tail of clients experiences significantly longer round-trip times with their DNS resolver.
For the solid, blue plot the 95th percentile and the 99th percentile have a value of 49~ms and 80~ms, respectively.

In total, our results indicate that our proposal significantly improves the performance of DNS resolutions even when the DNS resolver can provide the client with a cached response.
We find, that the long tail of clients experiencing high network latencies with their DNS resolver benefits the most from the proposed resolver-less DNS\@. 

\subsection{Expiration Time of DNS Records}

The retrieval of an average website requires about 20 connections to different hostnames~\cite{Sy2019_resumption_across}.
Thus, we assume that a web server supporting our proposal requires a fresh cache of about these 20 hostnames per website.
To approximate the number of required DNS lookups by the web server, we investigate in the following the expiration time of DNS records for popular websites.
We start by describing our methodology used to collect these DNS records, before we summarize our results.

\subsubsection{Data Collection}

For this measurement, we retrieve fresh DNS records directly from the authoritative nameserver of a given hostname.
In detail, we identified the authoritative nameserver of the Alexa Top Ten Thousand websites~\cite{Alexa} using the DNS lookup tool \textit{dig}.
Then, we used the provided nameserver to retrieve a fresh DNS record of the type A and AAAA for the corresponding hostname.
Subsequently, we retrieved the expiration time presented in the DNS record.
We conducted this measurement on the 5th of August 2019.
In total, we successfully retrieved the expiration time for 2401 AAAA and 9362 A DNS records.
The failures are mainly caused by nameservers not supporting the requested DNS record type and connection failures such as timeouts.

\subsubsection{Results}

\begin{figure}[t]
\centering
\includegraphics[width=0.60 \textwidth]{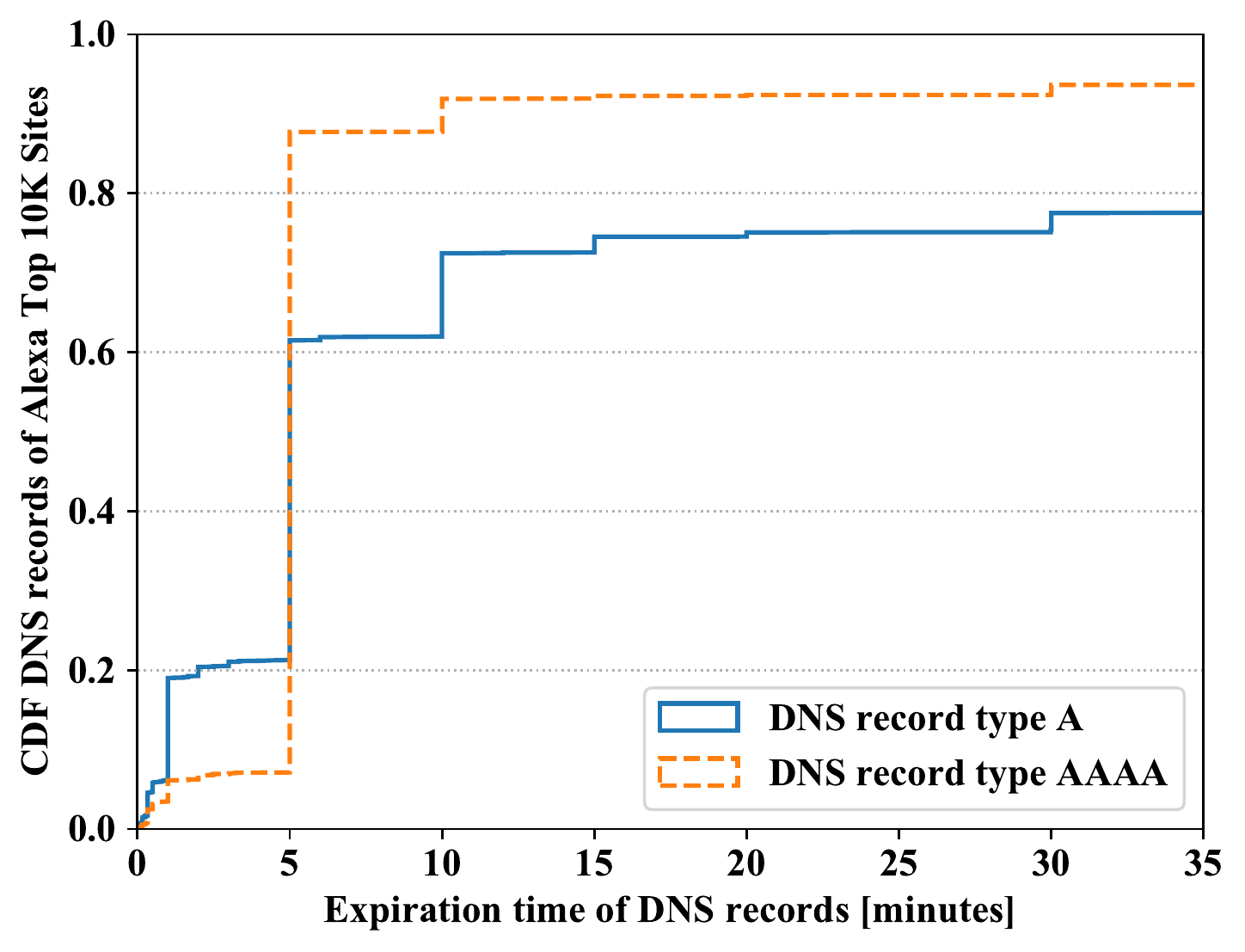}
\caption{Cumulative distribution of the collected DNS records of the Alexa Top 10K Sites over their expiration time.}
\label{fig:CDF_DNS_expiration}
%\vskip -12pt
\end{figure}

Figure~\ref{fig:CDF_DNS_expiration} plots a cumulative distribution of the retrieved DNS records of the Alexa Top Ten Thousands Sites over their expiration time.
The solid, blue and the dashed, orange plot represent the expiration times of A and AAAA DNS record types, respectively.
Our results indicate that an expiration time of five minutes is very popular within the investigated DNS records.
We find that 80.6\% of the AAAA and 40.2\% of the A DNS records use exactly this configuration.
Furthermore, A DNS records often use one and ten minutes expiration time with a share of 12.8\% and 10.5\%, respectively.
We observe, that only a tail of 12.5\% of the A records and 6.4\% of the AAAA records advertise expiration times longer than 30 minutes.
Moreover, our results indicate that 6.2\% of the A records and 3.5\% of the AAAA records announce an expiration time shorter than one minute.

In total, we find that resolver-less DNS is feasible with regard to real-world DNS record expirations times because they lead only to a small number of additional DNS lookups per minute per website. 

\subsection{ECS Support by Authoritative Nameserver}

EDNS client subnet intends to speed up the data delivery from CDNs by resolving DNS queries relative to the client's location as indicated by  an IP address prefix~\cite{rfc7871}.
However, ECS is a controversial performance optimization, especially with respect to its privacy properties~\cite{ECS_privacy}.
Furthermore, there exist reasonable alternatives such as anycast services~\cite{rfc4786} and the feature of connection migration within the QUIC transport protocol~\cite{Quic_socks}.
Nonetheless, we are investigating in this section the interaction between our proposal and real-word ECS\@.
We begin by describing the applied methodology of this measurement.
Subsequently, we summarize our results.
 
\subsubsection{Data Collection}

Google Public DNS is well-known to support ECS\@.
Thus, we retrieved the DNS records for the Alexa Top Ten Thousand Sites from Google Public DNS using the DNS lookup tool \textit{dig}.
\textit{Dig} allows us to explicitly signal support for ECS by specifying a client subnet.
During our measurements, we specified a client subnet matching the IP address of our test server located within a data center of our hosting provider Hetzner Online GmbH.
We conducted this measurement on the 5th of August 2019 and successfully retrieved 9999 DNS records that included an ECS response.
The failed retrieval can be attributed to a DNS failure of the authoritative nameserver.

\subsubsection{Results}

The analysis of the retrieved DNS records yields, that 88.8\% of these records announce a global scope.
Thus, only the remaining 11.2\% of the investigated hostnames provide DNS resolutions based on the presented client subnet.
The size of the scopes provided in these 11.2\% of the DNS records varied for our announced client subnet with the IPv4 address 159.69.184.0.
10.0\% of the responses covered a scope of /23 and /24 containing 512 and 256 IP addresses, respectively.
However, we assume that the results of this measurement vary depending on the announced client subnet.
Furthermore, we expect these scopes to be larger if the presented client's source address is part of a larger IP address block of the same AS as this may hint a physical proximity of these addresses.
Concerning our proposal, we prefer large scopes of IP addresses for which a DNS record is valid because this reduces the number of required DNS lookups by the web server.

In total, we find that almost 90\% of the DNS records are valid on a global scope which supports the feasibility of our proposal.
For the remaining hostnames ECS can make additional DNS lookups necessary.
However, this depends on the ECS configuration of the authoritative nameserver and the clients' source addresses.
%Total 9999 Responses 
%8884 ; CLIENT-SUBNET: 159.69.184.0/32/0
%  31 ; CLIENT-SUBNET: 159.69.184.0/32/13
%   1 ; CLIENT-SUBNET: 159.69.184.0/32/14
%  34 ; CLIENT-SUBNET: 159.69.184.0/32/16
%   1 ; CLIENT-SUBNET: 159.69.184.0/32/17
%  22 ; CLIENT-SUBNET: 159.69.184.0/32/20
%   6 ; CLIENT-SUBNET: 159.69.184.0/32/21
% 112 ; CLIENT-SUBNET: 159.69.184.0/32/23
% 888 ; CLIENT-SUBNET: 159.69.184.0/32/24
%  17 ; CLIENT-SUBNET: 159.69.184.0/32/32
%   4 ; CLIENT-SUBNET: 159.69.184.0/32/9

\section{Discussion} \label{sec:Discussion}

In this section, we compare the privacy and security properties of our proposal to the traditional DNS\@.
 We start by presenting our security considerations.
 Subsequently, we describe the privacy impact of the proposed resolver-less DNS\@.
 
\subsection{Security Considerations}

Translating the security threat of the resolver-less DNS to the traditional DNS, we have a scenario where the web server colludes with the client's recursive resolver.
Therefore, this scenario is similar to a client visiting google.com and using Google Public DNS as the recursive resolver.
In the following, we investigate the security risks of such a design by differentiating four cases:
\begin{enumerate}
\item The server presents a non-registered hostname and a fake DNS record.
\item The server presents a non-registered hostname and a true DNS record.
\item The server presents a registered hostname and a fake DNS record.
\item The server presents a registered hostname and the true DNS record.
\end{enumerate}
The process of domain name registration requires the registering entity to identify itself and therefore mitigates abuse.
In case 1 and 2, the adversary links to a non-registered hostname.
In case 1 the fake DNS record leads the client to connect to that non-registered hostname, while the true DNS record in case 2 prevents this connection establishment because there does not exist a DNS record for non-registered hostnames.
Thus, adversaries can exploit case 1 to launch for example phishing attacks with non-registered, look-alike hostnames aiming to steal user data/ passwords.
In case 3, the adversary can redirect traffic of a registered hostname to a malicious server using a fake DNS record.
However, there exist some mitigations such as the Strict-Transport-Security HTTP response header that can be used to prevent user agents to connect to these hostnames using an insecure connection.
Case 4 describes the designated use case where the client connects to the correct IP address of the registered hostname.

In total, only case 1 and 3 present security issues and they can always be mitigated when the established connections deploy server authentication as it is the default for TLS connections.
If the client attempts to establish insecure connections to a hostname using a DNS record retrieved via resolver-less DNS, it must make a second DNS lookup on this hostname using a fallback DNS mechanism.
Subsequently, the client must only send application data using this insecure connection if both DNS mechanisms yielded the same result.
Otherwise, the name resolution of the fallback DNS mechanism must be used to establish the insecure connection.
This approach prevents attacks derived from case 1 or 3, if the fallback DNS mechanism still provides trustworthy DNS records. 

The resilience against censorship via tampered DNS resolver presents another important security aspect.
Here, resolver-less DNS enables user agents to retrieve DNS records for domain names which cannot be correctly resolved by the client's tampered DNS resolver.
Thus, our proposal makes censorship on the web more difficult when these connections are established using server authentication.

Based on the above security considerations, we find that the security guarantees of our proposed resolver-less DNS are adequate or even better than the guarantees of the traditional DNS\@.

\subsection{Privacy Considerations}

The deployment of resolver-less DNS contributes to a reduced number of DNS queries a client has to send towards the used fallback DNS resolver.
This benefits the privacy posture of the user as the DNS resolver obtains only a diminished view on the user's browsing activities.
However, our proposal does not replace the traditional DNS and occasionally requires fallback DNS lookups.
For example, DNS queries to recursive resolvers are necessary to bootstrap the first connection which then can provide relevant DNS records using our proposal.
Furthermore, the proposed resolver-less DNS requires user agents to verify DNS records used for insecure connections via the traditional DNS\@.

To address these remaining privacy leaks, we proposed a whitelisting mechanism for DNS records from trusted websites possibly those hosted by Google or Cloudflare. 
As a result, these whitelisted DNS records are trusted to same extend as DNS records originating from the fallback DNS resolver.
Thus, the user agent is not required to conduct fallback DNS lookups for the purpose of validating the DNS records received via resolver-less DNS.
In total, we find a widespread deployment of resolver-less DNS would significantly improve the user's privacy posture with respect to its used recursive DNS resolver.

\section{Related Work} \label{sec:Related}
In this section, we compare our proposal to related work sharing the goal of a faster and/or a more privacy-friendly DNS\@.

Oblivous DNS~\cite{schmitt2019oblivious} is designed to prevent the recursive resolver from learning the user's IP address.
For this purpose, a second DNS resolver is used as a proxy that learns only the client's IP address but not the content of the DNS query.
This design provides significant privacy protections compared to the status quo.
As a drawback, Oblivous DNS increases the DNS lookup time due to the additional latency and computational overhead.
Furthermore, it does not support the performance-optimization EDNS client subnet.
In total, resolver-less DNS provides better performance properties than Oblivious DNS but is not always supported by web server.
Thus, Oblivious DNS may be a privacy-friendly alternative to the traditional DNS whenever our proposal requires to fetch DNS records via a fallback mechanism.

The DNS Anonymity Service combines a broadcast mechanism for popular DNS records with an anonymity network to conduct additional DNS lookups~\cite{federrath2011privacy}.
Unlike our proposal, the DNS Anonymity Service causes additional network traffic for downloading the broadcasted DNS records and suffers additional network latency when the client resolves hostnames via the anonymity network.
In total, the performance gains of this clean-slate approach are vague as they depend on the user's browsing behavior.
Furthermore, this approach does not integrate well into the existing DNS and requires additional Internet infrastructure to be deployed.

DNS prefetching describes a popular performance optimization where browsers start resolving the hostname of hyperlinks before the user clicks on them.
However, privacy research on this mechanism indicates severe privacy problems.
For example, it was shown that the recursive resolver could even infer the search terms the user entered into the search engine based on DNS prefetching~\cite{DNS-prefetching}.

\section{Conclusions}\label{sec:Conclusion}

As our objective, we hope to raise awareness for the performance and privacy limitations of traditional DNS\@.
To address these real-world problems, we presented resolver-less DNS\@.
Our evaluation substantiates the feasibility of this proposal and its significant performance and privacy benefits.
Furthermore, the proposed design is rather simple and operators of web server are incentivized to support resolver-less DNS as it improves the page loading time of their websites.

%%
%% Bibliography
%%

%% Please use bibtex, 

\bibliography{lipics-v2019-sample-article, sample-bibliography}

\begin{thebibliography}{10}

\bibitem{ietf-quic-http-22}
Mike Bishop.
\newblock {Hypertext Transfer Protocol Version 3 (HTTP/3)}.
\newblock Internet-Draft draft-ietf-quic-http-22, Internet Engineering Task
  Force, July 2019.
\newblock Work in Progress.
\newblock URL:
  \url{https://datatracker.ietf.org/doc/html/draft-ietf-quic-http-22}.

\bibitem{Privacy_by_Infrastructure}
Samantha Bradshaw and Laura DeNardis.
\newblock {Privacy by Infrastructure: The Unresolved Case of the Domain Name
  System}.
\newblock {\em Policy \& Internet}, 11(1):16--36, 2019.
\newblock URL: \url{https://onlinelibrary.wiley.com/doi/abs/10.1002/poi3.195},
  \href
  {http://arxiv.org/abs/https://onlinelibrary.wiley.com/doi/pdf/10.1002/poi3.195}
  {\path{arXiv:https://onlinelibrary.wiley.com/doi/pdf/10.1002/poi3.195}},
  \href {https://doi.org/10.1002/poi3.195} {\path{doi:10.1002/poi3.195}}.

\bibitem{callahan2013modern}
Thomas Callahan, Mark Allman, and Michael Rabinovich.
\newblock {On modern DNS behavior and properties}.
\newblock {\em ACM SIGCOMM Computer Communication Review}, 43(3):7--15, 2013.

\bibitem{chung2017longitudinal}
Taejoong Chung, Roland van Rijswijk-Deij, Balakrishnan Chandrasekaran, David
  Choffnes, Dave Levin, Bruce~M Maggs, Alan Mislove, and Christo Wilson.
\newblock {A Longitudinal, End-to-End View of the {DNSSEC} Ecosystem}.
\newblock In {\em 26th {USENIX} Security Symposium ({USENIX} Security 17)},
  pages 1307--1322, 2017.

\bibitem{rfc7871}
Carlo Contavalli, Wilmer van~der Gaast, David~C Lawrence, and Warren~"Ace"
  Kumari.
\newblock {Client Subnet in DNS Queries}.
\newblock RFC 7871, May 2016.
\newblock URL: \url{https://rfc-editor.org/rfc/rfc7871.txt}, \href
  {https://doi.org/10.17487/RFC7871} {\path{doi:10.17487/RFC7871}}.

\bibitem{federrath2011privacy}
Hannes Federrath, Karl-Peter Fuchs, Dominik Herrmann, and Christopher Piosecny.
\newblock {Privacy-preserving DNS: analysis of broadcast, range queries and
  mix-based protection methods}.
\newblock In {\em European Symposium on Research in Computer Security}, pages
  665--683. Springer, 2011.

\bibitem{Resource_Hints}
Ilya Grigorik.
\newblock {Resource Hints}.
\newblock {W3C} working draft, W3C, March 2019.
\newblock https://www.w3.org/TR/2019/WD-resource-hints-20190307/.

\bibitem{herrmann2013behavior}
Dominik Herrmann, Christian Banse, and Hannes Federrath.
\newblock {Behavior-based tracking: Exploiting characteristic patterns in DNS
  traffic}.
\newblock {\em Computers \& Security}, 39:17--33, 2013.

\bibitem{hounsel2019analyzing}
Austin Hounsel, Kevin Borgolte, Paul Schmitt, Jordan Holland, and Nick
  Feamster.
\newblock {Analyzing the costs (and benefits) of DNS, DoT, and DoH for the
  modern web}.
\newblock {\em arXiv preprint arXiv:1907.08089}, 2019.

\bibitem{HTTP_Archive}
{\relax HTTP Archive}.
\newblock {Report: State of the Web}, 2019.
\newblock URL: \url{https://www.httparchive.org/reports/state-of-the-web}.

\bibitem{Alexa}
Alexa~Internet Inc.
\newblock {Alexa Top 1,000,000 Sites}, 2019.
\newblock URL: \url{http://s3.amazonaws.com/alexa-static/top-1m.csv.zip}.

\bibitem{jung2002dns}
Jaeyeon Jung, Emil Sit, Hari Balakrishnan, and Robert Morris.
\newblock {DNS performance and the effectiveness of caching}.
\newblock {\em IEEE/ACM Transactions on networking}, 10(5):589--603, 2002.

\bibitem{kang2015my}
Ruogu Kang, Laura Dabbish, Nathaniel Fruchter, and Sara Kiesler.
\newblock “my data just goes everywhere:” user mental models of the
  internet and implications for privacy and security.
\newblock In {\em Eleventh Symposium On Usable Privacy and Security ({SOUPS}
  2015)}, pages 39--52, 2015.

\bibitem{ECS_privacy}
Panagiotis Kintis, Yacin Nadji, David Dagon, Michael Farrell, and Manos
  Antonakakis.
\newblock {Understanding the Privacy Implications of ECS}.
\newblock In Juan Caballero, Urko Zurutuza, and Ricardo~J. Rodr{\'i}guez,
  editors, {\em Detection of Intrusions and Malware, and Vulnerability
  Assessment}, pages 343--353, Cham, 2016. Springer International Publishing.

\bibitem{DNS-prefetching}
Srinivas Krishnan and Fabian Monrose.
\newblock {DNS Prefetching and Its Privacy Implications: When Good Things Go
  Bad}.
\newblock In {\em Proceedings of the 3rd USENIX Conference on Large-scale
  Exploits and Emergent Threats: Botnets, Spyware, Worms, and More}, LEET'10,
  pages 10--10, Berkeley, CA, USA, 2010. USENIX Association.
\newblock URL: \url{http://dl.acm.org/citation.cfm?id=1855686.1855696}.

\bibitem{rfc4786}
Kurt~Erik Lindqvist and Joe Abley.
\newblock {Operation of Anycast Services}.
\newblock RFC 4786, December 2006.
\newblock URL: \url{https://rfc-editor.org/rfc/rfc4786.txt}, \href
  {https://doi.org/10.17487/RFC4786} {\path{doi:10.17487/RFC4786}}.

\bibitem{rfc8288}
Mark Nottingham.
\newblock {Web Linking}.
\newblock RFC 8288, October 2017.
\newblock URL: \url{https://rfc-editor.org/rfc/rfc8288.txt}, \href
  {https://doi.org/10.17487/RFC8288} {\path{doi:10.17487/RFC8288}}.

\bibitem{pan2003overview}
Jianping Pan, Y~Thomas Hou, and Bo~Li.
\newblock {An overview of DNS-based server selections in content distribution
  networks}.
\newblock {\em Computer Networks}, 43(6):695--711, 2003.

\bibitem{pearce2017global}
Paul Pearce, Ben Jones, Frank Li, Roya Ensafi, Nick Feamster, Nick Weaver, and
  Vern Paxson.
\newblock {Global Measurement of {DNS} Manipulation}.
\newblock In {\em 26th {USENIX} Security Symposium ({USENIX} Security 17)},
  pages 307--323, 2017.

\bibitem{ripe}
{\relax Reseaux IP Europeens Network Coordination Centre}.
\newblock {RIPE Atlas -- Internet measurement network}, 2019.
\newblock URL: \url{https://atlas.ripe.net/}.

\bibitem{schmitt2019oblivious}
Paul Schmitt, Anne Edmundson, Allison Mankin, and Nick Feamster.
\newblock {Oblivious DNS: practical privacy for DNS queries}.
\newblock {\em Proceedings on Privacy Enhancing Technologies},
  2019(2):228--244, 2019.

\bibitem{velocity}
Steve Souders.
\newblock {Velocity and the Bottom Line}, 2009.
\newblock URL:
  \url{http://radar.oreilly.com/2009/07/velocity-making-your-site-fast.html/}.

\bibitem{Sy_TLS_Tracking}
Erik Sy, Christian Burkert, Hannes Federrath, and Mathias Fischer.
\newblock {Tracking Users Across the Web via TLS Session Resumption}.
\newblock In {\em Proceedings of the 34th Annual Computer Security Applications
  Conference}, ACSAC '18, pages 289--299, New York, NY, USA, 2018. ACM.
\newblock URL: \url{http://doi.acm.org/10.1145/3274694.3274708}, \href
  {https://doi.org/10.1145/3274694.3274708}
  {\path{doi:10.1145/3274694.3274708}}.

\bibitem{Sy2019_resumption_across}
Erik Sy, Moritz Moennich, Tobias Mueller, Hannes Federrath, and Mathias
  Fischer.
\newblock {Enhanced Performance for the encrypted Web through TLS Resumption
  across Hostnames}.
\newblock {\em CoRR}, abs/1902.02531, 2019.
\newblock URL: \url{http://arxiv.org/abs/1902.02531}, \href
  {http://arxiv.org/abs/1902.02531} {\path{arXiv:1902.02531}}.

\bibitem{2019QuicSocks}
Erik {Sy}, Tobias {Mueller}, Moritz {Moennich}, and Hannes {Federrath}.
\newblock {Accelerating QUIC's Connection Establishment on High-Latency Access
  Networks}.
\newblock {\em arXiv e-prints}, Jul 2019.
\newblock \href {http://arxiv.org/abs/1907.01291} {\path{arXiv:1907.01291}}.

\bibitem{Quic_socks}
Erik Sy, Tobias Mueller, Moritz Moennich, and Hannes Federrath.
\newblock {Accelerating QUIC's Connection Establishment on High-Latency Access
  Networks}.
\newblock {\em CoRR}, abs/1907.01291, 2019.
\newblock URL: \url{http://arxiv.org/abs/1907.01291}, \href
  {http://arxiv.org/abs/1907.01291} {\path{arXiv:1907.01291}}.

\bibitem{Varvello:2016:EPC:2999572.2999590}
Matteo Varvello, Jeremy Blackburn, David Naylor, and Konstantina Papagiannaki.
\newblock {EYEORG: A Platform For Crowdsourcing Web Quality Of Experience
  Measurements}.
\newblock CoNEXT '16, pages 399--412, New York, NY, USA, 2016. ACM.
\newblock \href {https://doi.org/10.1145/2999572.2999590}
  {\path{doi:10.1145/2999572.2999590}}.

\end{thebibliography}

\end{document}